\documentclass[twocolumn,aps,superscriptaddress,showpacs,preprintnumbers,amsmath,amssymb]{revtex4}
\usepackage{graphicx,color}
\usepackage{dcolumn}
\usepackage{colordvi} 
\usepackage{bm}
\begin{document}
\title{Weak value amplification in a shot-noise limited interferometer}
\author{Atsushi Nishizawa}
\email{anishi@yukawa.kyoto-u.ac.jp}
\affiliation{Yukawa Institute for Theoretical Physics, Kyoto University, 
Kyoto 606-8502, Japan}
\author{Kouji Nakamura}
\affiliation{TAMA project, Optical and Infrared Astronomy Division, National Astronomical Observatory of Japan, Mitaka, Tokyo 181-8588, Japan}
\author{Masa-Katsu Fujimoto}
\affiliation{TAMA project, Optical and Infrared Astronomy Division, National Astronomical Observatory of Japan, Mitaka, Tokyo 181-8588, Japan}
\date{\today}

\begin{abstract}
We study the weak-value amplification (WVA) in a phase measurement with an optical interferometer in which shot noise limits the sensitivity. We compute the signal and the shot noise including the full-order interaction terms of the WVA, and show that the shot-noise contribution to a phase shift in a pointer variable is always larger than the final variance of the pointer variable. This yields difference in estimating noise level up to a factor of 1.5. To clarify an advantage for practical uses of the WVA, we discuss signal-to-noise ratio and its optimization in the presence of the shot noise.
\end{abstract}

\maketitle

\section{Introduction}
The idea of weak-value amplification (WVA) was originally introduced by Aharonov, Albert, and Vaidman (AAV) in 1988 \cite{bib2} (see \cite{bib19,bib20} for a review). For a weak interaction between a system and a measuring device, they showed that the measurement results can be much larger than the eigenvalues of the observables by appropriately selecting initial and final states of the system. This theoretical prediction has been demonstrated in various experiments: the rotation of photon polarization \cite{bib3,bib4,bib5}, quantum box problem \cite{bib6}, the arrival time of a single photon \cite{bib7}, the spin Hall effect of light \cite{bib8}, the beam deflection and phase measurements in a Sagnac interferometer \cite{bib9,bib10,bib11,bib12}, and charge sensing \cite{bib13}.

In the original proposal of AAV \cite{bib2}, they considered the situation of the measurement in which the interaction between the system and the measuring device is so weak that the linear approximation with respect to the interaction strength is valid. If one specifies the initial and final states $|\psi_{i}\rangle$ and $|\psi_{f}\rangle$ of the system, which are called pre- and post-selected states, the outcome of the measurement, i.e., the shift in the pointer variable of the measuring device after the post-selection, becomes the so-called weak value
\begin{equation}
A_w \equiv \frac{\langle \psi_f | A |\psi_i \rangle}{\langle \psi_f | \psi_i \rangle} \;, 
\label{eq3}
\end{equation}
where $A$ is an observable associated with the system to be measured. If the outcome of the measurement is exactly this weak value, it seems that we have an arbitrarily large outcome when the pre- and post-selections are nearly orthogonal \cite{bib14}. However, when the weak value becomes large, non-linear effects of the von-Neumann measurement affect the outcome of the measurement \cite{bib15,bib16,bib17,bib1}. As the result, the shift of the pointer variable has a maximum value and vanishes when the pre- and post-selected states are exactly orthogonal. This means that there exists the optimal strength of the interaction and the optimal choice of the pre- and post-selected states.

Among the above application of the weak measurements, the effects measured in \cite{bib9,bib10} are an optical beam deflection transverse to the light propagation direction in a Sagnac interferometer. The authors have experimentally demonstrated that the small tilt of a mirror is actually amplified and detected. They also applied this technique to measurement of tiny phase shift \cite{bib11} and frequency stabilization \cite{bib12}.
These series of the experiments have shown that the WVA significantly improves signal-to-noise ratio (SNR) in the situation where technical noise (e.g. alignment noise) dominates \cite{bib10}. Also Brunner and Simon \cite{bib21} investigated the WVA for the measurement of a small longitudinal phase shift and concluded that WVA has potential to outperform a standard interferometry by 3 orders of magnitude in the presence of technical noise. However, as pointed out in \cite{bib10}, if photon shot noise dominates, no such a significant improvement of the SNR is achieved. Nevertheless, there is an interesting feature that the SNR is linearly proportional to the initial variance of the pointer variable of the measuring device. More recently, Parks and Gray \cite{bib18} discussed that the final variance of the pointer variable could be arbitrarily small if the pre-selected state is tuned so as to cancel the initial variance. However, from a practical point of view, the photon shot noise always exists and it is unclear whether the above small final variance of a pointer variable actually improves the SNR in real experiments. In addition, we note that these researches in \cite{bib10,bib18,bib21} are based on the linear approximation of the interaction strength and do not includes non-linear effects due to the von-Neumann interaction. Therefore, the improvement of the SNR and the advantage of the WVA are still ambiguous in an optical phase measurement, especially, in the situation where photon shot noise dominates.

In this paper, we study the phase measurement with the WVA in an optical interferometer, in which the phase shift is induced by small mirror displacement. Taking full-order evaluation of the von-Neumann interaction and the shot noise due to photon-number fluctuations into consideration, we derive the formula for the shot noise and discuss the optimization of the SNR. As a result, we found that the shot noise is always larger than the final variance of an pointer variable. We also discuss the detection limit of mirror displacement and the improvement of the sensitivity by WVA compared with a standard interferometry.   

This paper is organized as follows. In Sec.~\ref{sec2}, we explain the WVA and compute the expectation value of the $n$-th power of an pointer-variable, which can be applied to the arbitrary strength of the interaction and the arbitrary initial state of the pointer variable. In Sec.~\ref{sec3}, we derive the formula of shot noise and evaluate SNR in the presence of the shot noise. In Sec.~\ref{sec4}, based on the expression of SNR found in Sec.~\ref{sec3}, we discuss detection limit of mirror displacement. Sec.~\ref{sec5} is devoted to conclusions and discussions. Throughout this paper, we adopt the unit $c=\hbar=1$.

\section{Weak value amplification}
\label{sec2}

\begin{figure}[t]
\begin{center}
\includegraphics[width=8cm]{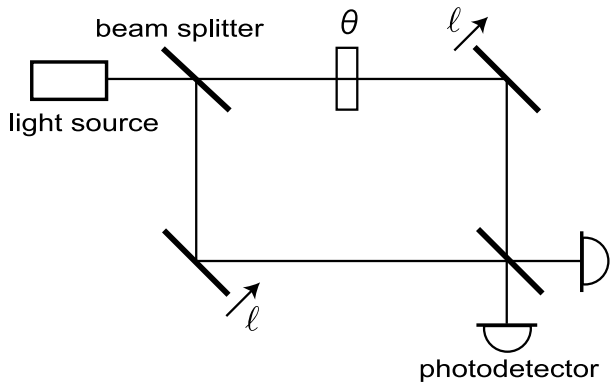}
\caption{Mach-Zehnder interferometer.}
\label{fig1}
\end{center}
\end{figure}

We consider a small displacement measurement of a mirror in an optical interferometer, perticularly concentrating on a Mach-Zehnder interferometer. This does not lose generality and our analysis here can also be applied to other optical configuration such as Michelson and Sagnac interferometers. The Mach-Zehnder interferometer we consider is shown in Fig.~\ref{fig1}. A light beam that enters the interferometer takes two paths after being divided by a 50/50 beam splitter. One of the beams on the upper path is phase-shifted by $\theta$ (pre-selection), which introduces a constant phase difference between beams on the upper and lower paths. Then each beam senses small differential displacements $\ell$ of mirrors (weak interaction) and is recombined at the second beam splitter. Finally, the light is detected at one of the output ports that is tuned to a nearly dark port by controling the initial phase difference $\theta$ (post-selection and sequential ideal measurement).

We present the theoretical description of WVA in this section for a single photon. However, the results can be easily generalized to a macroscopic beam. In our optical setup, we regard the beam which-path information as the system in weak measurement, which is the linear system spanned by the base states $\{|\uparrow \, \rangle$, $|\downarrow \,\rangle \}$. Here, $|\uparrow\rangle$ and $|\downarrow\rangle$ are the states that a photon propagates on the upper and lower optical paths of the interferometer, respectively. 
We regard the photon momentum as the pointer variable of the measuring device, which measures the phase shift induced by the mirror displacements in the Mach-Zehnder interferometer.

The pre-selected state of the system (photon propagation state) is denoted by 
\begin{equation}
| \psi_i \rangle = \frac{1}{\sqrt{2}} \left( e^{i \theta /2} | \uparrow \,\rangle + e^{-i \theta /2} |\downarrow \,\rangle \right) \;. \nonumber
\end{equation}
Here the initial phase offset $\theta$ is symmetrized merely for simplicity of calculation. The initial state of the measuring device (the probability distribution of photon momentum) is
\begin{equation}
| \Phi \rangle = \int dp\, \Phi(p) |p \rangle \;. \nonumber
\end{equation}
Since we measure the small displacement of the mirror at the asymmetric output port of this optical configuration, the observable is $\sqrt{2} \ell A$ where $A=| \uparrow \, \rangle \langle \, \uparrow |-| \downarrow \, \rangle \langle \, \downarrow | $. The Hamiltonian of the interaction at the time $t_0$ is written as  
\begin{equation}
H = g \delta(t-t_0) \, A \otimes p \;. 
\label{eq2}
\end{equation}
Here we defined $g \equiv \sqrt{2} \ell$. After the interaction given in Eq.~(\ref{eq2}) and the post-selection by the final state of the system $|\psi_f \rangle =( | \uparrow \, \rangle - | \downarrow \, \rangle )/\sqrt{2}$, the final state of the device is  
\begin{align}
|\Phi^{'} \rangle &= \langle \psi_f | e^{-i g A p} |\psi_i \rangle | \Phi \rangle \nonumber \\
&= \int dp \,\Phi (p) |p \rangle \langle \psi_f | e^{-i g A p} |\psi_i \rangle \;. \nonumber
\end{align}
Since the operator A satisfies the property $A^2=1$, this expression can be exactly evaluated including nonlinear terms in the coupling \cite{bib1} as
\begin{equation}
|\Phi^{'} \rangle = \int dp \,\Phi (p) |p \rangle \langle \psi_f |\psi_i \rangle (\cos gp -i A_w \sin gp) \;, 
\label{eq4}
\end{equation}
where $A_w$ is the weak value defined in Eq.~(\ref{eq3}) and is given in this case by
\begin{equation}
A_w = -i \cot \frac{\theta}{2} \;.
\label{eq6}
\end{equation}
Since the product of the initial and final states of the system is $\langle \psi_f | \psi_i \rangle = i \sin( \theta /2 )$, the state $|\Phi^{'} \rangle$ is not normalized. Here we define the density matrix of the measuring device 
\begin{equation}
\rho_d^{'} \equiv \frac{| \Phi^{'} \rangle \langle \Phi^{'} |}{\langle \Phi^{'} | \Phi^{'} \rangle} \;. 
\label{eq1}
\end{equation}
From Eqs.~(\ref{eq4}) and (\ref{eq1}), the expectation value of the $n$-th power of $p$ is 
\begin{align}
\langle p^n \rangle^{'} &= {\rm{Tr}}[ p^n \rho_d^{'} ] \nonumber \\
&= \frac{\langle p^n \rangle + (|A_w|^2-1) \langle p^n \sin^2 gp \rangle + {\rm{Im}} A_w \langle p^n \sin 2gp \rangle}{1+ (|A_w|^2-1) \langle \sin^2 gp \rangle + {\rm{Im}} A_w \langle \sin 2gp \rangle} \;. 
\label{eq5}
\end{align}
The bracket $\langle \cdots \rangle$ and $\langle \cdots \rangle^{'}$ denote averaging over the initial and final state of the measuring device, respectively.

If we measure the shift of the pointer variable $\langle p \rangle^{'}$, the variance ${\rm{Var}} [p]^{'} = \langle p^2 \rangle^{'} - (\langle p \rangle^{'})^2$ makes the pointer variable fluctuate and is regarded as frequency noise. However, in an optical experiment, shot noise coming from the fluctuation of photon number also contributes. In the next section, we will show that the shot noise is always larger than the frequency noise.

\section{Shot noise and SNR}
\label{sec3}
We consider the shot noise in an optical experiment of WVA and evaluate its contribution to the total noise. We note that the ingredients in Sec.~\ref{sec3a} are independent of the details of the interferometer setup and are applicable to other interferometers. We will also consider the optimization of the SNR. Hereafter we use frequency $\omega$ for photons instead of momentum $p$ in the previous section, though both variables denote the same quantity.

\subsection{Derivation}
\label{sec3a}
At the output port of the interferometer, we obtain a frequency spectrum of the photon number $n(\omega)$, where $\omega$ is frequency of a photon. To estimate the frequency shift of the photon-number distribution, we need to spectroscopically detect the output light with a multi-channel photodetector. For a single photon, the frequency shift is given by
\begin{equation}
\langle \omega \rangle^{'} = \int d\omega\, \omega \langle \omega | \rho_d^{'} | \omega \rangle \;, 
\nonumber
\end{equation}
where $\langle \omega | \rho_d^{'} | \omega \rangle$ is the probability distribution function of a photon after the post-selection and satisfy the normalization condition
\begin{equation}
\int d\omega \, \langle \omega | \rho_d^{'} | \omega \rangle =1 \;. \nonumber
\end{equation}
For $N_{\rm{out}}$ output photons, the averaged photon-number distribution can be written as
\begin{equation}
\overline{n (\omega)} = N_{\rm{out}} \langle \omega | \rho_d^{'} | \omega \rangle \;. \nonumber
\end{equation}
Here we assume that the output light in each frequency mode is a coherent state. The expectation value of the frequency shift is given by
\begin{equation}
\langle \omega \rangle^{'} = \frac{1}{N_{\rm{out}}} \int d\omega\, \omega \, \overline{n (\omega)} \;. 
\nonumber
\end{equation}

In a real experiment, since the photon number fluctuates around the expectation value as $n(\omega)= \overline{n(\omega)}+\Delta n (\omega)$, the observed frequency shift also fluctuates like
\begin{align}
\tilde{\omega} &= \frac{1}{N_{\rm{out}}} \int d\omega\, \omega n (\omega) \;, \nonumber \\
&= \langle \omega \rangle^{'} + \Delta \omega \;, \nonumber 
\end{align}
where
\begin{equation}
\Delta \omega \equiv \frac{1}{N_{\rm{out}}} \int d\omega\, \omega \Delta n (\omega) \;. 
\nonumber
\end{equation}
Here $\Delta \omega$ is shot noise due to photon-number fluctuations. Since the output in each frequency mode is a coherent state, the photon number fluctuates according to the Poisson distribution. We denote averaging over the Poisson distribution in each frequency mode by a bracket with the subscript ${\rm{P}}$. Then, the expectation value is $\langle \Delta \omega \rangle_{\rm{P}}=0$ by definition. The variance is
\begin{align}
{\rm{Var}} [ \Delta \omega ] &= \langle (\Delta \omega)^2 \rangle_{\rm{P}}  \nonumber \\
&= \frac{1}{N_{\rm{out}}^2} \int d\omega \int d\omega^{'}\, \omega \omega^{'} \langle \Delta n (\omega) \Delta n (\omega^{'}) \rangle_{\rm{P}} \nonumber \\
&=  \frac{1}{N_{\rm{out}}^2} \int d\omega\, \omega^2 \overline{n(\omega)}  \nonumber \\
&= \frac{1}{N_{\rm{out}}} \langle \omega^2 \rangle^{'} \;,
\end{align}
where we used mode independency
\begin{equation}
\langle \Delta n (\omega) \Delta n (\omega^{'}) \rangle_{\rm{P}}  = \overline{n (\omega)} \delta (\omega-\omega^{'}) \;.
\end{equation}
Therefore, the SNR is
\begin{equation}
{\rm{SNR}} = \frac{| \langle \omega \rangle^{'} |}{\sqrt{{\rm{Var}}[\Delta \omega]}} = \sqrt{N_{\rm{out}}} \frac{| \langle \omega \rangle^{'} |}{\sqrt{\langle \omega^2 \rangle^{'}}} \;,
\label{eq8}
\end{equation}
As this result shows, the shot noise is always larger than the frequency noise, $\sqrt{{\rm{Var}}[\omega]^{'}}=\sqrt{\langle \omega^2 \rangle^{'} -(\langle \omega \rangle^{'})^2}$, which is often regarded as a fundamental noise in weak measurement. This fact means that we have to take the shot noise into account when we evaluate the improvement of SNR by the WVA. 

In the derivation, we take the infinitesimal frequency bins for the convenience of the computation. However, it can be easily varified that dividing into finite bins also leads to the same result, though the frequency integrals in an expectation value is replaced with summations. 

\subsection{Evaluation}
To evaluate the dependence of the shot noise on the experimental parameters ($s$ defined below and $\theta$), we consider a single photon case and assume that the initial momentum distribution of a photon is non-zero-mean Gaussian (For multiple photons, a pulsed laser whose central frequency is mode-locked to $\omega_0$): 
\begin{equation}
\Phi  (\omega) = \left( \frac{1}{2\pi \sigma_{\omega}^2} \right)^{1/4} \exp \left[ -\frac{(\omega-\omega_0)^2}{4 \sigma_{\omega}^2} \right] \;, \nonumber
\end{equation}
Here $\sigma_{\omega}^2$ is the variance of $\omega-\omega_0$. For this Gaussian initial state, using Eq.~(\ref{eq5}), we obtain the first and second powers of $\omega-\omega_0$:
\begin{align}
g \langle \omega - \omega_0 \rangle^{'} &= \frac{s\, e^{-s} \left\{ (|A_w|^2-1) \sin \beta + 2\,{\rm{Im}}A_w \cos \beta  \right\} }{2 Z} \;, 
\label{eq9} \\
g^2 \langle (\omega- \omega_0)^2 \rangle^{'} &= \frac{s}{2} \biggl[ 1+ \frac{s\,e^{-s}}{Z} \nonumber \\
& \times \left\{ (|A_w|^2-1) \cos \beta -2\, {\rm{Im}} A_w\, \sin \beta \right\} \biggr] \;, \label{eq10}
\end{align}
where
\begin{align}
s &\equiv 2 g^2 \sigma_{\omega}^2 \;, \quad \quad \beta \equiv 2 g \omega_0 \;,\nonumber \\
Z &\equiv 1+\frac{1}{2} \left( |A_w|^2-1 \right) \left( 1-e^{-s}\,\cos \beta \right) + e^{-s} {\rm{Im}} A_w\, \sin \beta \;. \nonumber
\end{align}
The parameter $s$ can be interpreted as measurement strength, since large $g$ means strong coupling of the interaction and large $\sigma_{\omega}$ means that the narrow temporal distribution of photons. Equivalently, substituting the explicit expression of $A_w$ in Eq.~(\ref{eq6}), the above equations are written as
\begin{align}
g \langle \omega - \omega_0 \rangle^{'} &= -\frac{s\, e^{-s} \sin (\theta-\beta)}{1-e^{-s} \cos (\theta -\beta)} \;, 
\label{eq11} \\
g^2 \langle (\omega- \omega_0)^2 \rangle^{'} &= \frac{s}{2} \left[ 1+ \frac{2s\,e^{-s}\,\cos (\theta-\beta)}{1-e^{-s} \cos (\theta -\beta)} \right] 
\label{eq12} \;.
\end{align} 
In a typical experiment, $\theta$ is technically limited to above $\sim 10^{-3}$ and $\beta$ is $\sim \ell \,\omega_0 \sim \ell / \lambda_0$. If we focus on the sensitive experiment measuring a small phase shift, $\beta$ can be neglected and the initial distribution can be well approximated by zero-mean Gaussian. Therefore, hereafter we set $\beta=0$ or $\omega_0=0$.  

\begin{figure}[t]
\begin{center}
\includegraphics[width=8cm]{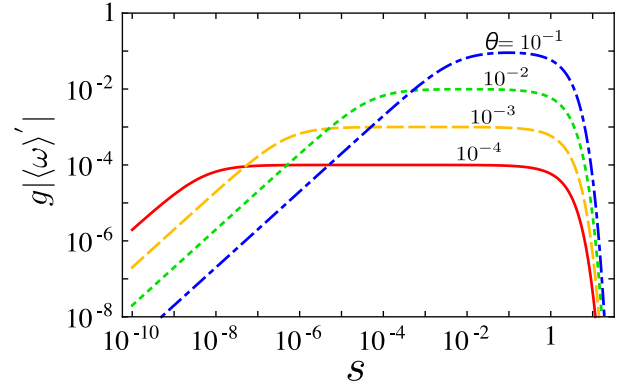}
\caption{Frequency shift $g \langle \omega \rangle^{'}$ as a function of $s$. Each curve is for $\theta=10^{-4}$ (red, solid), $10^{-3}$ (orange, dashed), $10^{-2}$ (green, dotted), and $10^{-1}$ (blue, dotted-dashed).}
\label{fig2}
\end{center}
\end{figure}

\begin{figure}[t]
\begin{center}
\includegraphics[width=8cm]{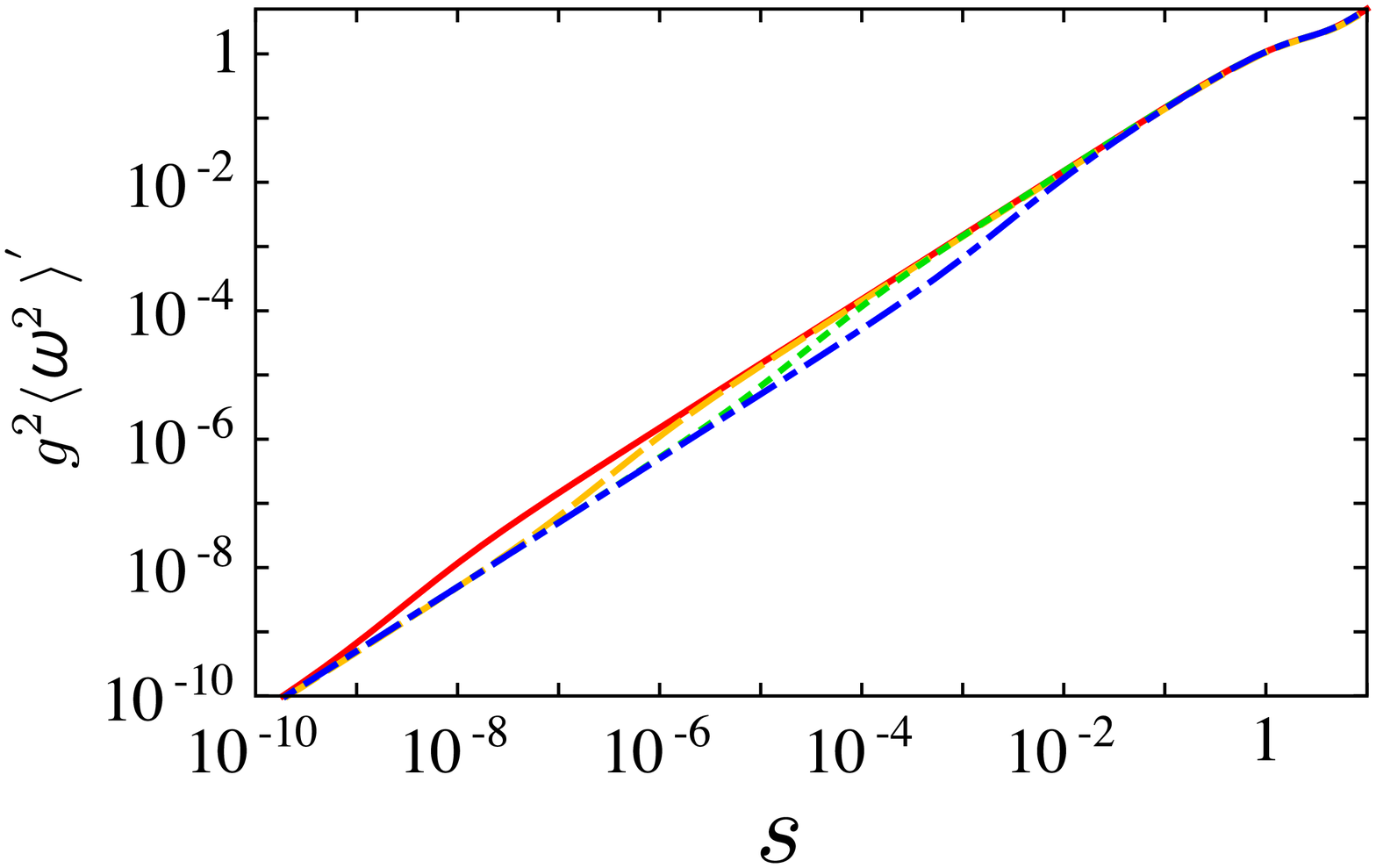}
\caption{Shot noise $g^2 \langle \omega^2 \rangle^{'}$ as a function of $s$. The curve types are the same as in Fig.~\ref{fig2}.}
\label{fig3}
\end{center}
\end{figure}

In Fig.~\ref{fig2}, the frequency shifts as a function of measurement strength $s$ for fixed $\theta$ are shown. For $s \ll1$, which is a weak measurement regime, the frequency shift linearly increases as $s$ is large. However, the amplification is saturated in an intermediate regime and rapidly drops in a strong measurement regime ($s \geq 1$). This behavior results from a nonlinear aspect of the WVA \cite{bib15,bib16,bib17,bib1}. Note that the maximum value of the frequency shift is larger as $\theta$ increases. In Fig.~\ref{fig3}, the shot noise as a function of measurement strength $s$ for fixed $\theta$ are shown. The magnitude of the shot noise linearly increases in broad range of the measurement strength. However, it hardly depends on $\theta$. These facts mean that main contribution of the shot noise comes from the initial variance of the measuring device, $s/2$. As is obvious from the definitions of the shot noise and the frequency noise (the final variance of the pointer variable), the shot noise coincides with the frequency noise for small values of the frequency shift. However, the shot noise deviate from the frequency noise at most by a factor of 1.5 for the Gaussian initial state. Note that the difference can be striking and qualitatively important for a non-Gaussian initial state because the frequency noise can be tuned to zero by appropriately selecting the initial state \cite{bib18}.

\begin{figure}[t]
\begin{center}
\includegraphics[width=8cm]{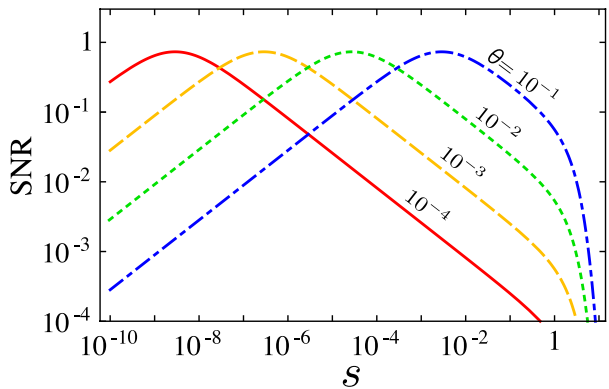}
\caption{SNR as a function of $s$. The curve types are the same as in Fig.~\ref{fig2}.}
\label{fig4}
\end{center}
\end{figure}

Although the frequency shift is actually amplified, the most important quantity in a real experiment is SNR. To see the dependence of the parameters, $s$ and $\theta$, we set $N_{\rm{out}}=1$ in Eq.~(\ref{eq8}). From Eqs.~(\ref{eq8}), (\ref{eq9}), and (\ref{eq10}), the SNR is given by
\begin{equation}
{\rm{SNR}}= \frac{\sqrt{2s} \,e^{-s} \sin \theta}{\sqrt{(1-e^{-s} \cos \theta) 
\{ 1-(1-2s) e^{-s} \cos \theta \} }} \;. 
\label{eq14}
\end{equation} 
In Fig.~\ref{fig4}, the SNRs as a function of measurement strength $s$ for fixed $\theta$ are shown. In contrast to the frequency shift, the SNR has a peak at a certain $s$. The optimal measurement strength is larger for a larger $\theta$. Unless the shot noise is properly taken into account, the SNR would be overestimated at most by a factor of 1.5 around the peaks of the SNR.

\subsection{Optimization of signal-to-noise ratio}
As seen in the previous subsection, the SNR has a maximum value for each fixed $\theta$. Next, we optimize the SNR and calculate the optimal $s$ and $\theta$. 

For a given $\theta$, the condition $\frac{\partial \,{\rm{SNR}}}{\partial s} =0$ is reduced to the equation to determine the optimal $s$ ($s_{\rm{opt}}$),
\begin{equation}
\cos \theta = e^{s_{\rm{opt}}} \left[ 1+s_{\rm{opt}}^2-s_{\rm{opt}} \left(1+\sqrt{3-2s_{\rm{opt}}+s_{\rm{opt}}^2} \right) \right] \;. 
\label{eq7}
\end{equation}
Figure~\ref{fig4} indicates that the SNR for smaller $\theta$ is optimized by smaller $s$. So for $s_{\rm{opt}} \ll 1$, Eq.~(\ref{eq7}) can be expanded in power of $s_{\rm{opt}}$ and gives 
\begin{equation}
s_{\rm{opt}} \approx \frac{1-\cos \theta}{\sqrt{3}} \approx \frac{\theta^2}{2\sqrt{3}} \;. \nonumber
\end{equation}
This relation well agrees with the locations of the SNR peaks in Fig.~\ref{fig4}. The maximum value of the SNR is obtained when $\theta \rightarrow 0$ and $s_{\rm{opt}} \rightarrow 0$ and is given by ${\rm{SNR}}_{\rm{max}}=\sqrt{2/(2+\sqrt{3})} \approx 0.732$. We note that this SNR is for a single output photon. In a typical optical experiment, $N_{\rm{out}}$ is significantly large and the SNR is improved proportional to $\sqrt{N_{\rm{out}}}$, giving the SNR much larger than unity.

\section{Detection limit of mirror displacement}
\label{sec4}
In this section, let us consider the detection limit of mirror displacement $\ell$, which is
estimated from the measurement strength parameter, $s=4 \ell^2 \sigma_{\omega}^2$. Since we found in the previous section that the SNR rapidly decreases in a strong measurement regime ($s \geq 1$), we concentrate on the weak regime and expand the SNR in powers of $s$. From Eq.~(\ref{eq14}) for $N_{\rm{out}}$ output photons, the leading contribution is
\begin{equation}
{\rm{SNR}} \approx \sqrt{2s N_{\rm{out}}} \cot \frac{\theta}{2} \;. \nonumber
\end{equation}
The output photon number $N_{\rm{out}}$ has $\theta$ dependence due to the choise of pre- and post-selection states and is significantly suppressed, compared to initial photon number $N_{\rm{in}}$, as $N_{\rm{out}}=N_{\rm{in}} \sin^2 (\theta/2)$. Setting ${\rm{SNR}}=1$ and solving for $\ell$, we obtain the detection limit of $\ell$:
\begin{equation}
\ell_{\rm{min}} = \frac{\lambda_0 }{4\pi \sqrt{2N_{\rm{in}}} \cos (\theta/2)} \left( \frac{\sigma_{\omega}}{\omega_0} \right)^{-1}  \;, \nonumber
\end{equation}
where $\lambda_0$ is the wavelength corresponding to $\omega_0$. 

Currently available lasers can generate femtosecond pulses. For example, if the parameters are selected as $\theta=10^{-3}$, $\lambda_0=1\,\mu {\rm{m}}$, $\sigma_{\omega}/\omega_0 =1$, and $N_{\rm{in}}=1\,{\rm{J}}/(\hbar \omega_0) \approx 5 \times 10^{18}$, the detection limit is $\ell_{\rm{min}}\approx 1.5 \times 10^{-17}\,{\rm{m}}$. For $1\,{\rm{sec}}$ of the integration time, this is comparable sensitivity with the measurement with a continuous monochromatic laser whose power is $1\,{\rm{W}}$ except for the factor $(\sigma_{\omega}/\omega_0)^{-1}$ \cite{bib22}.

\section{Conclusions and discussions}
\label{sec5}
We have studied a phase measurement with the WVA in an optical (Mach-Zehnder) interferometer, taking shot noise due to photon-number fluctuations into account. In this optical configuration, we have derived the formula of the WVA, namely the expectation value of the $n$-th power of a pointer-variable after post-selection. This formula can be applied to the arbitrary strength of the interaction and the arbitrary initial state of the pointer variable. Furthermore, we also derived the formula for the shot noise and discussed the optimization of the SNR. As a result, we found that the shot noise is always larger than the frequency noise. Also we showed that the SNR with smaller fixed $\theta$ is optimized by smaller $s$ ($s_{\rm{opt}} \sim \theta^2$). Although our results are shown by setting $N_{\rm{out}}=1$, the SNR is considerably improved due to a large number of photons, i.e. ${\rm{SNR}} \propto \sqrt{N_{\rm{out}}}$. 

The detection limit we derived is comparable to the standard measurement scheme with a continuous monochromatic laser except for the factor $(\sigma_{\omega}/\omega_0)^{-1}$. This conclusion is consistent with that claimed by Starling et al. \cite{bib10} in the measurement of a transverse beam deflection with a Sagnac interferometer. We emphasize that our estimation of the photon shot noise is general and is applicable to many interferometer setups of phase measurements, because the shot noise formula includes the non-linear effects of the von-Neumann interaction and does not depends on a specific optical configuration of the interferometer. 

Finally, we comment on one of the possibilities of the further improvement of SNR. Although we restrict the initial state of a probe to Gaussian distribution, further improvement of SNR would be possible by the introduction of the non-Gaussian initial state of the measuring device \cite{bib18,bib23}. Even if we apply non-Gaussian initial states, we have to take the photon shot noise into account and the arguments developed in this paper will be necessary. To clarify the advantage of practical applications of the WVA, further detailed and technical investigations are necessary.

\begin{acknowledgments}
The authors would like to thanks to A. Hosoya, Y. Shikano, K. Somiya and Y. Susa for valuable discussions and comments. A. N. is supported by a Grant-in-Aid through JSPS.
\end{acknowledgments}

\bibliography{wv-paper}

\end{document}